\documentclass[12pt,a4paper,twoside,final,notitlepage, leqno]{article}
\usepackage[english]{babel}
\usepackage[T1]{fontenc}  
\usepackage{graphicx}
\usepackage{amsmath,amsthm,epsfig,amsfonts,bbm}  

\newcommand{\pequationdeb}{$$ \left\{ \begin{minipage}[c]{130mm}}
\newcommand{\pequationfin}{\end{minipage}
                           \right. $$}

\def \smb {{\scriptstyle \bullet }}
 
\newcommand{\moneq}{\vspace*{-6pt} \begin{equation} \displaystyle } 
\newcommand{\moneqstar}{\vspace*{-6pt} \begin{equation*} \displaystyle } 
\newcommand{\monendstar}{\vspace*{-6pt} \end{equation*}   }
\newcommand{\monend}{\vspace*{-6pt} \end{equation}   }
\newcommand{\beq}     {\begin{equation}}
\newcommand{\enq}     {\end{equation}}
\newcommand{\be}    {\begin{enumerate}}
\newcommand{\ee}    {\end{enumerate}}

\newcommand{\Bb}

\def\R{{\rm I}\! {\rm R}}

\def\tvi {\vrule  height 10pt depth 5pt width 0pt}
\def\tv  {\tvi \vrule}
\def\tvg {\tv ~~}
\def\tvd {~~ \tv}
\def \na{ \noalign {\hrule}  }
\def\hcr {\hfill & \cr}

\def\br {\break}

\def\section*#1{}

\def\resume{\if@twocolumn
\section*{R\'esum\'e}
\else \small
\quotation{\bf \it R\'esum\'e \rule[1mm]{1.5mm}{0.2mm}\vspace{0pt}}
\fi}
\def\endresume{\if@twocolumn\else\endquotation\fi}
\def\abstract{\if@twocolumn
\noindent\section*{{\bf Abstract}}
\else \small
\quotation{\noindent \bf {Abstract.} \rule[1mm]{1.5mm}{0.2mm}\vspace{0pt}}
\fi}
\def\endabstract{\if@twocolumn\else\endquotation\fi}

\usepackage{fancyhdr}
\renewcommand{\headrulewidth}{0pt}
\begin{document}

\fancypagestyle{plain}{ \fancyfoot{} \renewcommand{\footrulewidth}{0pt}}
\fancypagestyle{plain}{ \fancyhead{} \renewcommand{\headrulewidth}{0pt}} 





~

\bigskip \bigskip   \bigskip \bigskip   \bigskip \bigskip

\centerline {\bf \Large  A quantum approach}
 
\smallskip \smallskip

\centerline {\bf \Large  
 for determining   a  state of the opinion}  

\bigskip \bigskip 
\centerline { \large  Fran\c{c}ois Dubois}

\centerline { \it  \small  
  Conservatoire National des Arts et M\'etiers  } 

\centerline { \it  \small   
  Department of Mathematics,   Paris, France. } 


\centerline { \it \small  francois.dubois@cnam.fr}  
 
 \bigskip
\centerline {  09  September 2012~\protect\footnote{~This contribution
 has been presented on Thuesday 03 July 2012 
at  the session ``Quantum Decision Theory'' during the symposium
 {\it Foundations and Applications of Utility, Risk and Decision Theory} (FUR), 
Georgia State University, Atlanta, USA. Edition 04 January 2013.}}   

\bigskip 
\noindent  {\bf Abstract } \qquad 
 We propose to define a notion of state
of the opinion in order to link politician popularity estimations 
and voting intentions. We present two ways of modelling: a classical approach and quantum
modelling. We test these  ideas on data obtained during the French
presidential election of April 2012. 

\smallskip 
\noindent  {\bf   Keywords}\qquad    Opinion polls, voting.  
  
  \bigskip  
  
\bigskip  \bigskip  \noindent {\bf \large 1) \quad   Introduction } 

\smallskip   \quad   
Electoral periods are favorable to opinion polls. 
We keep in mind that 
opinion polls  are intrinsically complex
(see {\it e.g.}  Gallup \cite{Ga44} or  Till\'e \cite {Ti01}) 
and give an approximates picture of a possible social reality. 
They are traditionnally of two types: popularity polls  for various 
outstanding political personnalities 
and  voting intentions  polls    when a list of candidates is known.
%
We remark that in the first case, a grid of appreciation 
is given by the questionnaire, typically of the type 
\quad  ``very  good''  $\, \succ \,$ 
``good''        $\, \succ \,$  
``no opinion''      $\, \succ \,$  
``bad''      $\, \succ \,$  
``very bad''.

  \smallskip \noindent $\bullet$ \quad  
We have   two different informations and to construct a link between them is not an easy task. 
In particular, the determination of the voting intentions is a quasi intractable  problem!
Predictions of votes classically use   of so-called ``voting functions''.
Voting functions  have been developed for the
prediction of presidential elections in the United States. 
They are based on correlations 
between economical parameters, popularity polls and other technical parameters. 
%
%
We refer to  Abramowitz \cite {Ab88}, Lewis-Beck \cite {LB91},  
Campbell  \cite{Ca92}, Lafay \cite{La97} and the survey paper proposed by 
Auberger \cite{Au04}.

  \smallskip \newpage 
\noindent $\bullet$ \quad  
We do not detail here  the mathematical difficulties associated with the question 
of voting when the number of candidates is greater than  three 
\cite  { Ar51, Bo1781, Co1785}. 
They conduct to present-day researches like range voting,  
  independently proposed by Balinski and Laraki \cite{BL07a, BL07b} 
and  by Rivest and Smith   \cite {Sm2k, RS07}. 
%
 It is composed by two steps: grading and ranking. In the  grading step, 
all the candidates are evaluated by all the electors. 
This first step is quite analogous to a popularity  investigations and 
we will merge the two notions in this contribution. 
The second step of range voting is a  majority ranking;
it consists of a successive extraction of medians. 

  \smallskip \noindent $\bullet$ \quad  
In this contribution, we make the hypothesis that there exists some global 
state of the opinion associated with   a 
given grid of analysis, denoted by  ${\rm G}$ 
in the following.  
We study how voting intentions interact with 
the state of the opinion. 
%
In particular, we propose 
to determine as much information as possible about this
state of the opinion, in the case where voting intentions 
and popularity polls  are both available. 
In Section~2, we propose a mathematical 
model  founded on a classical framework. The state of the opinion
is described by a discrete law of probability and the double information
of popularity polls  and opinion polls  give the input  information. 
 
  \smallskip \noindent $\bullet$ \quad  
We adopt  afterwards in Section~3 quantum modelling 
(see {\it e.g.} Bitbol {\it et al} \cite{Bi09} for an introduction), 
in the spirit of authors like   
 Khrennikov and Haven \cite{KH07}, La Mura and  Swiatczak \cite{LMS07} 
and Zorn and  Smith \cite{ZS11} concerning voting processes.   
We recall  two voting models developed in previous contributions  
\cite{Du08,  Du09}, founded on  
range voting and first run of an election, having  implicitely in mind  
the case of the French presidential election. 
Then we propose  in Section~4 to link our two quantum models and use for doing this
an equivalent candidate and the state of the opinion. 
%
We  test in Section~5  our previous ideas with three  sets of data coming from 
2012 French presidential elections and propose numerical  results.

\fancyfoot[C]{\oldstylenums{\thepage}}
\fancyhead[OC]{\sc{A quantum approach  for determining a state of the opinion}} 

\bigskip  \bigskip  
 \noindent {\bf \large 2) \quad   A classical approach } 

\smallskip   \quad   
We consider a grid $\rm G$ of $m$ types of opinions as one of the two following ones.  
We have $ \, m=5 \,$ for the first grid (\ref{grille-5}) and  
$ \, m=3 \,$ for the second one  (\ref{grille-3}):   
\moneq  \label{grille-5}
 ++  \,\, \,\succ \, \,\, 
 +  \,\,\, \succ \,\,\,
 0  \,\,\, \succ \,\,\, 
 -  \,\,\, \succ \,\,\,
 -- 
\monend 
\vskip -.7  cm
\moneq  \label{grille-3} 
 +  \,\,\, \succ \,\,\,
 0  \,\,\, \succ \,\,\, 
 -   \,. 
\monend 
These ordered grids are typically  used for popularity polls 
\cite {If12a, If12b, Ip12, Ip09avr12}.
We assume also that a ranking grid like 
(\ref{grille-5}) or (\ref{grille-3}) is a basic tool  
to represent a ``state of the opinion''. 
If some political personality has a great proportion of ``very good'' 
or ``$++$''  opinion (as in  (\ref{grille-5})), 
we suppose here that this fact is a kind of mirror effect
of an existing state of social opinion. 
The reflection that the opinion 
is for a certain proportion in a  ``very good'' state.  

  \smallskip \noindent $\bullet$ \quad  
We have  two type of data, as explained in the introduction. 
We denote by $\, \Gamma \, $ the set of candidates and we denote by $n$ 
their number. 
We suppose also that    
\moneq  \label{n-gt-m}  
{\rm the} \,{\rm number} \,  {\rm of} \,  {\rm candidates}   \,\,\,  \equiv \,\,\, n \,\, > \,\,   
m     \,\,\,  \equiv \,\,\, {\rm the} \, \, {\rm size} \, {\rm of}  \, {\rm the} \, 
 {\rm  grid}  \, \, {\rm G} \, .    
\monend 
On one side  the result of a popularity poll for the $n$ candidates 
is given. 
We have a matrix of data $ \, \big( S_{\gamma \, \nu} \big)_{\gamma \in \Gamma, \, \nu \in G}  \,$ 
with an hypothesis of coherence: 
\moneq  \label{matrice-S}
 S_{\gamma \, \nu} \geq 0 \,, \quad \sum_{\nu \in G}  S_{\gamma \, \nu} = 1 \,, \quad
 \gamma \in \Gamma \, .     
\monend 
On the other side, we have the voting intentions 
  $ \,  \beta_{\gamma}  \,$ for each candidate $ \, \gamma \in \Gamma.   \,$ 
We have at our disposal a vector  
  $ \, \beta \, \equiv \, \big( \beta_{\gamma} \big)_{\gamma \in \Gamma}  \,$ 
with $n$ components and satisfying 
\moneqstar 
  \beta_{\gamma}  \geq 0 \,, \quad \sum_{\gamma \in \Gamma}  \beta_{\gamma} \, \leq \, 1
  \, , \quad \gamma \in \Gamma \, . 
\monendstar  
In other words, 
\moneqstar 
\beta \, \in \, \widetilde{K}_n \, \equiv \, \Big\{ q \in \R^n , \,\, q_j \geq 0 \, , \,\, 
\sum_{j=1}^n  q_j \,\leq\, 1 \, \Big\} \, .  
\monendstar  

  \smallskip \noindent $\bullet$ \quad  
We adopt in this section a classical  point of view for taking into account the variety of possibles
underlyings. 
We suppose that the opinion $ \, \nu \,$ (with $\, \nu \in  {\rm G}$) 
is present in the entire population with a probability $ \,   p_\nu .\,$ 
So the  state of opinion is mathematically modelized by a   law of probability 
$ \, \big( p_\nu \big)_{\nu \in {\rm G}} .\,$   
The state of the opinion $\, p \, $ satisfies the  
 natural constraints 
\moneq  \label{contrainte-p}
p \, \in \, K_m \, \equiv \, \Big\{ q \in \R^m , \,\, q_j \geq 0 \, , \,\, 
\sum_{j=1}^m  q_j \,=\, 1 \, \Big\} \,  
\monend 
that express that we have a discrete law of probability. 
There are two natural questions when we try to link the vector
$ \, \beta \, $ of voting intentions with the state of opinions $ \,p$.

\smallskip \noindent {\bf (Q$_{\bf 1}$)}  \quad If the state of the opinion is known, 
how to predict the voting
intentions ?

\smallskip \noindent  {\bf (Q$_{\bf 2}$)} \quad If the voting intentions are known, how to determine
  the    state of the opinion  ?  
 
  \smallskip \noindent $\bullet$ \quad  
The answer to the question  (Q$_1$) is simple if we consider that voting intentions 
could be determined by the state of the opinion. Then 
we think coherent to express that the expectation of the family 
 $ \, S_{\gamma \, \nu} \,$
for $ \, \nu \,$ running in $\rm G$ is equal to the voting intention  
$ \,  \beta \in  \widetilde{K}_n  .  \,$
We can say also that the correlation of the probability vectors $p$
and $ \, s_{\gamma} \equiv  \big( S_{\gamma \, \nu} \big)_{\nu \in {\rm G}}  \,$ 
is equal to the voting intention $\, \beta_\gamma .\,$ 
In algebraic terms, 
\moneq  \label{esperance} 
\sum_{\nu \in G}  S_{\gamma \, \nu} \,\, p_\nu \,=\, \beta_{\gamma} \,,\quad \gamma \in
\Gamma  \, .    
\monend 
The question  (Q$_2$) exchanges the datum and the unknown. Then the relation 
(\ref{esperance}) is now a linear system with unknown $\, p \in  K_m \,$ 
and given datum $ \,  \beta \in  \widetilde{K}_n  .  \,$ 
Of course, the system (\ref{esperance}) is   in general not correctly posed if 
the hypothesis (\ref{n-gt-m}) is satisfied. We have $n$ equations 
and only $m$ unknowns. 
We adopt in this contribution a least square approach and  replace the system  (\ref{esperance}) 
by the minimization of some squared functional, say 
\moneqstar 
J(p) \,=\, {1\over2} \sum_{\gamma \in \Gamma} 
\Big( \sum_{\nu \in G}  S_{\gamma \, \nu} \, p_\nu \,-\, \beta_{\gamma} \Big)^2  
\monendstar 
to fix the ideas. 
The constraint (\ref{contrainte-p}) has to be  satisfied because  
 the family of numbers   $ \, \big( p_\nu \big)_{\nu \in G} \,$  
is a probability distribution. 
We   solve a  quadratic optimization problem with the functional 
$\, J(\smb) \,$ and the  linear inequalities constraints (\ref{contrainte-p}):
\moneq  \label{pb-opti} 
\displaystyle 
{\rm find} \,\,\, p \, \in \, K_m \,\,\, {\rm such} \,\, {\rm that} \,\,\, 
J(p) \,=\, {\rm inf} \, \,  \big\{ J(q) ,\, \, q \in   K_m  \, \big\}   \, .    
\monend 
%
If the matrix  $ \, S_{\gamma \, \nu} \,$  introduced at the relation (\ref{matrice-S})
is of maximal rank $m$ (and we do this hypothesis in the following), 
the problem (\ref{pb-opti}) is the minimization of a coercive quadratic
functional inside  a closed non empty convex set. This problem has a unique solution;
we   solve it using the Uzawa algorithm (see {\it e.g.} the book
of Gondran and Minoux \cite{GM94}). 

\bigskip \bigskip  
 \noindent {\bf \large 3) \quad   Two quantum models for voting process } 

\smallskip   \quad  
The fact of  considering  quantum modelling induces a specific vision of probabilities. 
We refer {\it e.g.} to the classical  treatise  on quantum mechanics 
of  Cohen-Tannoudji {\it et al.} \cite{CDL77}, 
to the so-called contextual objectivity proposed by Grangier  \cite{Gr02},   
to the  approach of Mugur-Sch\"achter  \cite{MMS08},   
or to the elementary introduction 
proposed by Busemeyer and  Trueblood \cite  {BT09}
in the context of statistical inference.

  \smallskip \noindent $\bullet$ \quad  
In a  first tentative \cite{Du08}, we have proposed to introduce an 
Hilbert space  $\, V_\Gamma \,$ formally generated by the candidates 
$ \, \gamma \in \Gamma . \, $ In this space, a canditate $ \, \gamma \,$
is represented by a unitary vector $ \,   |  \, \gamma \! >  \,$ 
and this family of $n$ vectors is supposed to be orthogonal. 
Then an elector $ \, \ell \,$ can be decomposed 
in the space   $\, V_\Gamma \,$ of candidates according to
\moneq  \label{decompo-ell} 
\displaystyle 
 |  \, \ell \! > \,\,=\, \,
\sum_{\gamma \in \Gamma} \, \theta_{\ell  \gamma}  \,\,  | \, \gamma \! >  \, .     
\monend 
The vector $ \,  |  \, \ell \! > \,  \in V_\Gamma \,$ is supposed also to be 
a unitary vector to fix the ideas.  
According to Born's rule, the probability for a given elector 
$\, \ell \, $ to give his  voice to the particular candidate
$ \, \gamma \, $ is equal to $ \,  |  \,   \theta_{\ell  \gamma}  \,  | ^2 . \, $
The violence of the quantum measure is clearly visible  with this example: 
the opinions of an elector $ \, \ell \,$ never coincidate with the
program of any candidate. But with a voting system where an elector  has to choice 
only one candidate among $n$, his social opinion is {\it reduced} 
to the one of a  particular candidate.

  \smallskip \noindent $\bullet$ \quad  
Our  second model \cite{Du09} is adapted to  the grading step 
of range voting \cite{BL07a,Sm2k}.  
We introduce a specific grading space $ \, W_{\rm G} \,$ of political appreciations 
 associated with a grading
family $\rm G$. The space  $ \,  W_{\rm G} \,$ is formally generated by the $m$ orthogonal
vectors  $ \,  |  \, \nu \! > \,$ relative to the opinions. 
Then we suppose that the candidates $ \, \gamma \,$ are now decomposed 
by each elector 
on the basis   $ \,  |  \, \nu \! > : \,$
\moneq  \label{decompo-gamma} 
\displaystyle 
   |  \, \gamma \! > \,=\, \sum_{\nu  \in G} \, \alpha_{\nu \, \gamma} \, \, 
   |  \, \nu \! > \, , \quad \gamma \in \Gamma \, .  
 \monend 
Moreover the vector  $ \,  |  \, \gamma \! >  \,$ in   (\ref{decompo-gamma})
is supposed to be by a unitary:  
\moneq  \label{gamma-unitaire} 
\displaystyle 
 \sum_{\nu  \in G} \,   |  \,  \alpha_{\nu \, \gamma}  \,  | ^2  \,\, = \,\, 1 \, ,
\quad \gamma \in \Gamma \, . 
 \monend 
With this notation,  
the probability for a given elector  
to give an opinion 
$ \, \nu \,$ to a candidate $ \, \gamma \,$ is simply a consequence of the Born  rule.
The mean statistical
expectation   of a given opinion $\, \nu \,$ for a candidate  $ \, \gamma \,$ is equal to  
 $ \,  |  \,  \alpha_{\gamma \, \nu}  \,  | ^2  \, $  on one hand and is given by  
the popularity polls  $ \,  S_{\gamma \, \nu}  \,  $  on the other hand. Consequently, 
\moneq  \label{alpha-deux-egal-s} 
  |  \,  \alpha_{\nu \, \gamma}  \,  | ^2 \,=\,  S_{\gamma \, \nu} 
\,, \quad \gamma \in \Gamma \,, \,\, \nu \in {\rm G} . 
 \monend   
%

\bigskip \bigskip   
 \noindent {\bf \large 4) \quad   State of the opinion: a link between quantum 
 voting models }  

\smallskip \quad 
We have at our disposal two quantum models. The first one operates in an 
Hilbert space  $ \, V_\Gamma \,$ generated (formally) by the 
candidates $ \,  |  \, \gamma \! > \,$ for $ \, \gamma \in  \Gamma . \,$
The second uses an Hilbert space $ \, W_{\rm G} \,$ formally generated by the
grading $\rm G$ of appreciations 
$ \,  |  \, \nu \! > \,$ for $ \, \nu \in  {\rm G} . \,$

\smallskip \noindent $\bullet$ \quad 
The first model in space $ \, V_\Gamma \,$ 
is well adapted for determining  the voting intentions throught 
the Born rule. In this contribution, we simplify   the approach  (\ref{decompo-ell})
and suppose that there exists some equivalent candidate 
$ \,  |  \, \xi \! > \, \in V_\Gamma \,$ such that the voting intention
$ \, \beta_\gamma \,$ for each particular candidate $ \, \gamma \in  \Gamma  \,$
is equal to $\,  | <\xi \,,\, \gamma >  |^2 $:
\moneq  \label{equivalent-candidate} 
 | <\xi \,,\, \gamma >  |^2 \,=\,  \beta_\gamma \,, \quad \forall  \, \gamma \in  \Gamma 
\,; \quad  |  \, \xi \! > \, \equiv \, \sum_{\gamma \in \Gamma}  |  \, \gamma \! >
\,  <\gamma \,,\, \xi  > \, \,\, \in V_\Gamma \, . 
 \monend 
  
  \smallskip \noindent $\bullet$ \quad  
The second model in space $ \,  W_{\rm G} \, $ 
is appropriate to range voting and popularity polls. 
We interpret now the relation (\ref{decompo-gamma}) in the following way: for each
candidate $ \, \gamma \in \Gamma, \,$
there exists a political decomposition  $ \, A  \, |  \, \gamma \! > \, \in W_{\rm G} \,$
in terms of the grid G and we have 
\moneq  \label{opinion-gamma} 
\displaystyle  
A  \, |  \, \gamma \! >  \,=\, \sum_{\nu  \in G} \, \alpha_{\nu \, \gamma} \, \, 
   |  \, \nu \! > \, , \quad \gamma \in \Gamma \, .  
 \monend 
By linearity, we   construct in this way  a linear operator 
$\, A :\, V_\Gamma \longrightarrow W_{\rm G} \,$ between two different Hilbert spaces. 
A state of the opinion is now modelized by a vector 
$ \,  |  \, \zeta  \! > \, \in  W_{\rm G} .\,$ 
Remark that the coefficients $ \, \alpha_{\nu \, \gamma} \, $ are related
to the data $ \, S_{\gamma \, \nu} \,$ with the help of the relation 
(\ref{alpha-deux-egal-s}).
We suppose also $ \,  \alpha_{\nu \, \gamma} \geq 0 \,$ in the following
to fix the ideas.

  \smallskip \noindent $\bullet$ \quad 
The questions 
(Q$_1$) and (Q$_2$) presented in Section~2 can now be formulated in terms of links
between the equivalent candidate $ \,  |  \, \xi \! > \, \in V_\Gamma \,$ 
and the state of the opinion $ \,  |  \, \zeta  \! > \, \in  W_{\rm G} .\,$ 
If the state of the opinion  $ \,  |  \, \zeta  \! > \,  $ is known, 
the question set by (Q$_1$) is now to determine the voting intentions
 $\,  \beta_\gamma \, $ 
obtained also by the relation (\ref{equivalent-candidate}). 
We suppose that this operation is also  done
for each particular candidate $ \, \gamma \in  \Gamma  \,$
according to the Born rule {\it via} a scalar product between the opinion state 
 $ \,  |  \, \zeta  \! > \,  $ 
and the political decomposition  $ \, A  \, |  \, \gamma \! > \, $ 
proposed in  (\ref{opinion-gamma}): 
%
\moneq  \label{condition-raccord} 
\displaystyle  
 | <\zeta \,,\, A \, \gamma >  |^2  \,=\,  | <\xi \,,\, \gamma >  |^2 \,, \quad 
 \quad \forall \, \gamma \in  \Gamma  \,; \quad  
|  \, \zeta  \! > \, \in  W_{\rm G} \,. 
 \monend 
Then there exists some phase $ \, \varphi_\gamma \in \R \,$
for each $ \, \gamma \in \Gamma \,$ 
and the relation (\ref{condition-raccord}) implies 
\moneq  \label{condition-raccord-2}  
\displaystyle  
 <\zeta \,,\, A \, \gamma >   \,=\, {\rm e}^{\displaystyle   - i \, \varphi_\gamma} \, 
 <\xi \,,\, \gamma >   \,, \quad 
 \quad \forall  \, \gamma \in  \Gamma  \, . 
 \monend
We introduce the ``phase operator'' $\, J \,$ with a diagonal matrix
composed by the different phases:
\moneqstar 
\displaystyle  J \,=\, {\rm diag} \, \Big\{  {\rm e}^{\displaystyle    i \, \varphi_\gamma} 
\,,\, \gamma \in \Gamma \Big\} \, . 
 \monendstar 
We remark that the   adjoint operator $ \, J^* \,$ is the inverse 
 $ \, J^{-1} \,$ of the operator $J$:  $ \, J^* = J^{-1} .\,  $  
We introduce also the adjoint 
operator $ \, A^* : W_{\rm G} \longrightarrow V_\gamma .\,$
Then the relation  (\ref{condition-raccord-2})
takes the form 
\moneq  \label{condition-raccord-3}  
\displaystyle  
 <  A^* \,   \zeta \,,\, \gamma >   \,=\,  <\xi \,,\, J^* \, \gamma >  
 \, \equiv \,  < J \, \xi \,,\, \gamma >  
\,, \quad   \quad \forall  \, \gamma \in  \Gamma  \, . 
 \monend
By linearity of the operators $A$ and $J$, 
we can write the relation 
 (\ref{condition-raccord-3}) under the compact form 
\moneq  \label{condition-raccord-4}  
\displaystyle  
   A^* \,   |  \, \zeta  \! >     \,=\,   J \,   |  \, \xi   \! >   \,  . 
 \monend
We have a response to the first question (Q$_1$):
if the state of the opinion $ \,  |  \, \zeta  \! > \,  $  is known, 
it determines an equivalent candidate  $ \,  |  \, \xi   \! > \,  $ 
{\it modulo} a phase. We observe also that the phase operator is eliminated when we consider
the Born rule  (\ref{condition-raccord}). In the following, we 
replace the operator $ \, J \,$ by the identity and  (\ref{condition-raccord-4})
by 
\moneq  \label{condition-raccord-5}  
\displaystyle   
   A^* \,   |  \, \zeta  \! >     \,=\,      |  \, \xi   \! >   \,  . 
 \monend

  \smallskip \noindent $\bullet$ \quad 
The question  (Q$_2$) can now be formulated in a   simple way:
if the  equivalent candidate  $ \,  |  \, \xi   \! > \,  $ 
is known, is it possible to determine a state of the opinion 
 $ \,  |  \, \zeta  \! > \, \in W_{\rm G} \,  $
such that the relation  (\ref{condition-raccord-5}) holds ? 
The difficulty concerns now linear algebra. Because 
$ \, {\rm rank} \, A = m , \, $
the operator $ \, A^* \,$ is injective 
$ \,  W_{\rm G} \longrightarrow V_\Gamma .\, $
But it is not a surjective operator since $ \, n > m \,$
as supposed in  (\ref{n-gt-m}). 
In this contribution, we propose to solve   (\ref{condition-raccord-5})
in terms of least squares, {\it i.e.} to solve the equation obtained after 
multiplying the relation   (\ref{condition-raccord-5}) by the operator~$A$:
\moneq  \label{condition-raccord-6}  
\displaystyle  
   A \, A^* \,\,    |  \, \zeta  \! >      \,=\,  A \,\,       |  \, \xi \! >  \,  . 
 \monend
Then the state of the opinion  $ \,  |  \, \zeta  \! > \, $
can be determined without difficulty. We normalize it for our
application. 
When the state  $ \,  |  \, \zeta  \! > \, $ is known,  
 the relative quantum   probability $ \, \delta_\nu \,$ 
of observing the particular state $ \, \nu  \in {\rm G} \,$ 
  is equal, as  consequence of Born's rule,
 to the square of the component $ \,  < \nu ,\, \zeta >   \,$: 
  \moneq  \label{lien-theta-beta} 
 \delta_\nu \,\, = \,\,   
\mid     < \nu ,\, \zeta >   \mid  ^2  \,, \quad \nu  \in {\rm G}  \, . 
 \monend 
%

\bigskip    \bigskip    \bigskip 
 \noindent {\bf \large 5) \quad   Spring 2012 preliminary results } 

\smallskip   \quad  
We have obtained in French popular newspapers three political popularity polls in 
february, march and april 2012. 
For each case, we have chosen voting intention polls at 
a date as close as possible to the previous ones. 
The first family of data has been obtained in february 2012. 
Popularity data  \cite{If12a, Ip12}
and result of voting intentions  \cite{If12a, Ip12}
are displayed in Table~1. The names of the principal candidates 
to the French presidential election 
  are proposed in alphabetic order with the following  abbreviations:
``Ba'' for  Fran\c cois Bayrou,  
``Ho'' for  Fran\c cois Hollande, 
``Jo'' for  Eva Joly, 
``LP'' for  Marine Le Pen, 
``M\'e'' for  Jean-Luc M\'elanchon and 
``Sa'' for  Nicolas Sarkozy. 
Similar data are displayed in Table~2 for march 2012  \cite  {If12b, IF12d}
and  in Table~3 for april  2012  \cite{Ip09avr12,Ip10avr12}.
In this last table, we have also reported the result of the election 
of 22 April 2012.

%
%

   \bigskip

\setbox20=\hbox{ $\,\,$ }
\setbox30=\hbox{ $  + + $   }
\setbox40=\hbox{ $  + $  } 
\setbox50=\hbox{  ~ 0 } 
\setbox60=\hbox{ $  - $  } 
\setbox70=\hbox{ $ - - $ } 
\setbox80=\hbox{ voting } 
\setbox21=\hbox{ Ba }
\setbox22=\hbox{ Ho } 
\setbox23=\hbox{ Jo } 
\setbox24=\hbox{ LP } 
\setbox25=\hbox{ M\'e }
\setbox26=\hbox{ Sa } 
\setbox41=\hbox{ .55  }  
\setbox51=\hbox{ .14  }  
\setbox61=\hbox{ .31  }  
\setbox81=\hbox{ .125  }  
\setbox42=\hbox{ .52  }  
\setbox52=\hbox{ .08  }  
\setbox62=\hbox{ .40  }  
\setbox82=\hbox{ .30  }  
\setbox43=\hbox{ .29  }  
\setbox53=\hbox{ .13  }  
\setbox63=\hbox{ .58  }  
\setbox83=\hbox{ .03  }  
\setbox44=\hbox{ .28  }  
\setbox54=\hbox{ .06  }  
\setbox64=\hbox{ .66  }  
\setbox84=\hbox{ .175  }  
\setbox45=\hbox{ .38  }  
\setbox55=\hbox{ .20  }  
\setbox65=\hbox{ .42  }  
\setbox85=\hbox{ .085 }  
\setbox46=\hbox{ .33  }  
\setbox56=\hbox{ .00  }  
\setbox66=\hbox{ .67  }  
\setbox86=\hbox{ .25  }  
\setbox44=\vbox{\offinterlineskip  \halign {
&\tvg#& # &\tvg#&  #  &\tvg#& #  &\tvg#&  #&\tvd#&   #&\tvg#&  # &\tvd#\cr   
\na&   \box20  &&  \box40  &&  \box50  && \box 60 &&$\!\!$&&  \box80  \hcr  
\na&   \box21  &&  \box41  &&  \box51  && \box 61 &&$\!\!$&&  \box81  \hcr  
\na&   \box22  &&  \box42  &&  \box52  && \box 62 &&$\!\!$&&   \box82 \hcr   
\na&   \box23  &&  \box43  &&  \box53  && \box 63 &&$\!\!$&&   \box83 \hcr   
\na&   \box24  &&  \box44  &&  \box54  && \box 64 &&$\!\!$&&  \box84  \hcr   
\na&   \box25  &&  \box45  &&  \box55  && \box 65 &&$\!\!$&&  \box85  \hcr   
\na&   \box26  &&  \box46  &&  \box56  && \box 66 &&$\!\!$&&  \box86  
\hfill   \hcr  \na}   }  \centerline{\box44  }
%
\centerline {  {\bf  Table  1}.   \quad 
Popularity and sounding polls, february 2012 \cite{If12a,  If12c, Ip12}. }  
\bigskip

\newpage 

\setbox20=\hbox{ $\,\,$ }
\setbox30=\hbox{ $  + + $   }
\setbox40=\hbox{ $  + $  } 
\setbox50=\hbox{  ~ 0 } 
\setbox60=\hbox{ $  - $  } 
\setbox70=\hbox{ $ - - $ } 
\setbox80=\hbox{ voting } 
\setbox21=\hbox{ Ba }
\setbox22=\hbox{ Ho } 
\setbox23=\hbox{ Jo } 
\setbox24=\hbox{ LP } 
\setbox25=\hbox{ M\'e }
\setbox26=\hbox{ Sa } 
\setbox31=\hbox{ .08 }   
\setbox41=\hbox{ .62  }  
\setbox51=\hbox{ .03  }  
\setbox61=\hbox{ .23  }  
\setbox71=\hbox{ .04  }  
\setbox81=\hbox{ .12  }  
\setbox32=\hbox{ .09 }   
\setbox42=\hbox{ .45  }  
\setbox52=\hbox{ .00  }  
\setbox62=\hbox{ .30  }  
\setbox72=\hbox{ .16  }  
\setbox82=\hbox{ .275 }  
\setbox33=\hbox{ .02 }   
\setbox43=\hbox{ .34  }  
\setbox53=\hbox{ .02  }  
\setbox63=\hbox{ .40  }  
\setbox73=\hbox{ .22  }  
\setbox83=\hbox{ .03  }  
\setbox34=\hbox{ .10 }   
\setbox44=\hbox{ .24  }  
\setbox54=\hbox{ .01  }  
\setbox64=\hbox{ .26  }  
\setbox74=\hbox{ .39  }  
\setbox84=\hbox{ .17  }  
\setbox35=\hbox{ .11 }   
\setbox45=\hbox{ .46  }  
\setbox55=\hbox{ .03  }  
\setbox65=\hbox{ .31  }  
\setbox75=\hbox{ .09  }  
\setbox85=\hbox{ .11  }  
\setbox36=\hbox{ .10 }   
\setbox46=\hbox{ .31  }  
\setbox56=\hbox{ .00  }  
\setbox66=\hbox{ .29  }  
\setbox76=\hbox{ .30  }  
\setbox86=\hbox{ .28  }  
\setbox44=\vbox{\offinterlineskip  \halign {
&\tvg#& # &\tvg#&  #  &\tvg#&  #  &\tvg#&  #  &\tvg#&  #  &\tvg#&  #&\tvd#&   #&\tvg#&  # &\tvd#\cr    
\na&   \box20  && \box30 &&  \box40  &&  \box50  && \box 60 &&  \box70   &&$\!\!$&&   \box80 \hcr  
\na&   \box21  && \box31 &&  \box41  &&  \box51  && \box 61 &&  \box71   &&$\!\!$&&   \box81 \hcr  
\na&   \box22  && \box32 &&  \box42  &&  \box52  && \box 62 &&  \box72   &&$\!\!$&&   \box82  \hcr   
\na&   \box23  && \box33 &&  \box43  &&  \box53  && \box 63 &&  \box73   &&$\!\!$&&   \box83  \hcr   
\na&   \box24  && \box34 &&  \box44  &&  \box54  && \box 64 &&  \box74   &&$\!\!$&&   \box84  \hcr   
\na&   \box25  && \box35 &&  \box45  &&  \box55  && \box 65 &&  \box75   &&$\!\!$&&   \box85  \hcr   
\na&   \box26  && \box36 &&  \box46  &&  \box56  && \box 66 &&  \box76   &&$\!\!$&&   \box86  
\hfill   \hcr  \na}   }  \centerline{\box44  }
\smallskip \centerline {  {\bf  Table  2}.   \quad 
Popularity and sounding polls, march 2012 \cite  {If12b, IF12d}. }  
\bigskip

\bigskip 
\setbox20=\hbox{ $\,\,$ }
\setbox30=\hbox{ $  + +  $   }
\setbox40=\hbox{ $  +   $  } 
\setbox50=\hbox{  ~ 0  } 
\setbox60=\hbox{ $  -    $    } 
\setbox70=\hbox{ $ - -   $ } 
\setbox80=\hbox{ voting } 
\setbox90=\hbox{ result }  
\setbox21=\hbox{ Ba }
\setbox22=\hbox{ Ho } 
\setbox23=\hbox{ Jo } 
\setbox24=\hbox{ LP } 
\setbox25=\hbox{ M\'e }
\setbox26=\hbox{ Sa } 
\setbox41=\hbox{ .56  }  
\setbox51=\hbox{ .07  }  
\setbox61=\hbox{ .37  }  
\setbox81=\hbox{ .095  }  
\setbox91=\hbox{ .091  }  
\setbox42=\hbox{ .57  }  
\setbox52=\hbox{ .03  }  
\setbox62=\hbox{ .40  }  
\setbox82=\hbox{ .285 }  
\setbox92=\hbox{ .286 }  
\setbox43=\hbox{ .35  }  
\setbox53=\hbox{ .10  }  
\setbox63=\hbox{ .55  }  
\setbox83=\hbox{ .015 }  
\setbox93=\hbox{ .023 }  
\setbox44=\hbox{ .26  }  
\setbox54=\hbox{ .05  }  
\setbox64=\hbox{ .69  }  
\setbox84=\hbox{ .15  }  
\setbox94=\hbox{ .179 }  
\setbox45=\hbox{ .47  }  
\setbox55=\hbox{ .10  }  
\setbox65=\hbox{ .43  }  
\setbox85=\hbox{ .145 }  
\setbox95=\hbox{ .111 }  
\setbox46=\hbox{ .49  }  
\setbox56=\hbox{ .05  }  
\setbox66=\hbox{ .46  }  
\setbox86=\hbox{ .29  }  
\setbox96=\hbox{ .272 }  
\setbox44=\vbox{\offinterlineskip  \halign { 
&\tvg#& # &\tvg#&  #  &\tvg#& #  &\tvg#&  #&\tvd#&   #&\tvg#&  #&\tvd#&   #&\tvg#&  #&\tvd#\cr   
\na&   \box20  &&  \box40  &&  \box50  && \box 60 && $\!\!$&&  \box80 && $\!\!$&&  \box90  \hcr  
\na&   \box21  &&  \box41  &&  \box51  && \box 61 && $\!\!$&&  \box81 && $\!\!$&&  \box91  \hcr  
\na&   \box22  &&  \box42  &&  \box52  && \box 62 && $\!\!$&&  \box82 && $\!\!$&&  \box92  \hcr   
\na&   \box23  &&  \box43  &&  \box53  && \box 63 && $\!\!$&&  \box83 && $\!\!$&&  \box93  \hcr   
\na&   \box24  &&  \box44  &&  \box54  && \box 64 && $\!\!$&&  \box84 && $\!\!$&&  \box94  \hcr   
\na&   \box25  &&  \box45  &&  \box55  && \box 65 && $\!\!$&&  \box85 && $\!\!$&&  \box95  \hcr   
\na&   \box26  &&  \box46  &&  \box56  && \box 66 && $\!\!$&&  \box86 && $\!\!$&&  \box96  
\hfill   \hcr  \na}   }  \centerline{\box44  }
\smallskip \centerline {  {\bf  Table  3}.   \quad 
 Popularity, sounding polls and result,   april  2012 \cite{Ip09avr12,Ip10avr12}. } 
\bigskip

  \bigskip   
\smallskip \noindent $\bullet$ \quad 
The result  of our mathematical treatment is presented in tables 4 to 7. 
From popularity polls and   voting intentions, we evaluate a classical and a quantum 
state of the opinion. 
The first line gives the classical probability $ \, p \, $ solution of the problem 
(\ref{pb-opti}). The second line describes the components of the quantum state 
of the opinion $ \, \zeta \,$ compatible with the relation 
 (\ref{condition-raccord-6}). 
The third line is the quantum probability, {\it id est} 
the square of the second line according to Born's rule (see  (\ref{lien-theta-beta})).
We observe that the constraints (\ref{contrainte-p}) are active and induce 
values equal to zero for some classical probabilities
in the first line  of Table~5 and Table~6. The quantum state  $ \, \zeta \,$ (second 
line of tables~4 to~7) is unitary.
We observe that the sign of some components is negative. 
The  comparisons of classical and quantum probabilities (first and third lines
of tables~4 to~7) agree globally in a first approach.
Nevertheless for precise components (opinion ``$-$'' in march 2012 {\it id est}  
fourth column of Table~5, opinion
``+'' in april 2012 {\it id est} first  column of tables~6 and~7) 
the two probabilities  differ notabily. 

 \bigskip 
  
\bigskip  
\setbox20=\hbox{ $\,\,$ } 
\setbox40=\hbox{ $  + $  } 
\setbox50=\hbox{  ~ 0 } 
\setbox60=\hbox{ $  - $  }  
\setbox21=\hbox{ classical probability }
\setbox22=\hbox{ quantum state } 
\setbox23=\hbox{ quantum  probability }  
\setbox41=\hbox{ .15  }  
\setbox51=\hbox{ .74  }  
\setbox61=\hbox{ .11  }  
\setbox42=\hbox{ .34   }  
\setbox52=\hbox{ $-$.90  }  
\setbox62=\hbox{ .26   }  
\setbox43=\hbox{ .11  }  
\setbox53=\hbox{ .82  }  
\setbox63=\hbox{ .07  }  
\setbox44=\vbox{\offinterlineskip  \halign {
&\tvg#& # &\tvg#&  #  &\tvg#& #  &\tvg#&   # &\tvd#\cr   
\na&   \box20  &&  \box40  &&  \box50  && \box 60  \hcr  
\na&   \box21  &&  \box41  &&  \box51  && \box 61  \hcr   
\na&   \box22  &&  \box42  &&  \box52  && \box 62  \hcr 
\na&   \box23  &&  \box43  &&  \box53  && \box 63  
\hfill   \hcr  \na}   }  \centerline{\box44  }
\smallskip \centerline {  {\bf  Table  4}.   \quad 
Classical and quantum state of the opinion,  february 2012.  }  
\bigskip      

\newpage 
  
\setbox20=\hbox{ $\,\,$ }
\setbox30=\hbox{ $  + + $   }
\setbox40=\hbox{ $  + $  } 
\setbox50=\hbox{ ~  0 } 
\setbox60=\hbox{ $  - $  } 
\setbox70=\hbox{ $ - - $ }  
\setbox21=\hbox{ classical probability }
\setbox22=\hbox{ quantum state } 
\setbox23=\hbox{ quantum  probability }  
\setbox31=\hbox{ .57  }  
\setbox41=\hbox{ .14  }  
\setbox51=\hbox{  ~ 0 }  
\setbox61=\hbox{  ~ 0 }  
\setbox71=\hbox{  .29 }  
\setbox32=\hbox{ $-$.58 }  
\setbox42=\hbox{ .53  }  
\setbox52=\hbox{ $-$.18 }  
\setbox62=\hbox{ $-$.51   }  
\setbox72=\hbox{  .31 }  
\setbox33=\hbox{ .33  }  
\setbox43=\hbox{ .28  }  
\setbox53=\hbox{ .03  }  
\setbox63=\hbox{ .26  }  
\setbox73=\hbox{ .10  }  
\setbox44=\vbox{\offinterlineskip  \halign {
&\tvg#& # &\tvg#&  #  &\tvg#& #  &\tvg#&  #  &\tvg#&   #  &\tvg#&   # &\tvd#\cr   
\na&   \box20  &&  \box30  &&  \box40  &&  \box50  && \box 60  && \box 70  \hcr  
\na&   \box21  &&  \box31  &&  \box41  &&  \box51  && \box 61  && \box 71  \hcr   
\na&   \box22  &&  \box32  &&  \box42  &&  \box52  && \box 62  && \box 72  \hcr 
\na&   \box23  &&  \box33  &&  \box43  &&  \box53  && \box 63  && \box 73  
\hfill   \hcr  \na}   }  \centerline{\box44  }
\smallskip \centerline {  {\bf  Table  5}.   \quad 
Classical and quantum state of the opinion,  march 2012.  }  
\bigskip

\setbox20=\hbox{ $\,\,$ } 
\setbox40=\hbox{ $  + $  } 
\setbox50=\hbox{  ~ 0 } 
\setbox60=\hbox{ $  - $  }  
\setbox21=\hbox{ classical probability }
\setbox22=\hbox{ quantum state } 
\setbox23=\hbox{ quantum  probability }  
\setbox41=\hbox{ .28  }  
\setbox51=\hbox{ .72  }  
\setbox61=\hbox{ ~  0   }  
\setbox42=\hbox{ .14   }  
\setbox52=\hbox{ $-$.96  }  
\setbox62=\hbox{ .25   }  
\setbox43=\hbox{ .02  }  
\setbox53=\hbox{ .92  }  
\setbox63=\hbox{ .06  }  
\setbox44=\vbox{\offinterlineskip  \halign {
&\tvg#& # &\tvg#&  #  &\tvg#& #  &\tvg#&   # &\tvd#\cr   
\na&   \box20  &&  \box40  &&  \box50  && \box 60  \hcr  
\na&   \box21  &&  \box41  &&  \box51  && \box 61  \hcr   
\na&   \box22  &&  \box42  &&  \box52  && \box 62  \hcr 
\na&   \box23  &&  \box43  &&  \box53  && \box 63  
\hfill   \hcr  \na}   }  \centerline{\box44  }
\smallskip  \centerline {    {\bf  Table  6}.   \quad 
Classical and quantum state of the opinion,  april  2012.   }
\bigskip      

\setbox20=\hbox{ $\,\,$ } 
\setbox40=\hbox{ $  + $  } 
\setbox50=\hbox{  ~ 0 } 
\setbox60=\hbox{ $  - $  }  
\setbox21=\hbox{ classical probability }
\setbox22=\hbox{ quantum state } 
\setbox23=\hbox{ quantum  probability }  
\setbox41=\hbox{ .25  }  
\setbox51=\hbox{ .73  }  
\setbox61=\hbox{ .02  }  
\setbox42=\hbox{ .13   }  
\setbox52=\hbox{ $-$.96  }  
\setbox62=\hbox{ .25   }  
\setbox43=\hbox{ .02  }  
\setbox53=\hbox{ .92  }  
\setbox63=\hbox{ .06  }  
\setbox44=\vbox{\offinterlineskip  \halign {
&\tvg#& # &\tvg#&  #  &\tvg#& #  &\tvg#&   # &\tvd#\cr   
\na&   \box20  &&  \box40  &&  \box50  && \box 60  \hcr  
\na&   \box21  &&  \box41  &&  \box51  && \box 61  \hcr   
\na&   \box22  &&  \box42  &&  \box52  && \box 62  \hcr 
\na&   \box23  &&  \box43  &&  \box53  && \box 63  
\hfill   \hcr  \na}   }  \centerline{\box44  }
\smallskip  \centerline {   {\bf  Table  7}.   \quad 
Similar to Table 6, but the voting polls  have been replaced }

\centerline {  by the result of 22 April  
(last column of Table 3).   }
\bigskip


\bigskip \bigskip    
 \noindent {\bf \large 6) \quad    Conclusion and perspectives  }  

\smallskip \quad 
In this contribution, we have introduced a state of the opinion to analyse 
with a given degree of precision the variety of appreciations  of political programs. 
In a classical approach the state of the opinion is a discrete law of probability.
With quantum modelling, this state is a  vector in an Hilbert space of political appreciations. 
Two questions has been formulated. 
On one hand, how the knowledge of the state of the opinion determines 
the voting intentions ? 
The reverse question on the other hand: how the knowledge of voting
intentions can define a state of the opinion~?
We have studied these two questions in both classical and quantum points of view.
We  have proposed  responses as simple as possible in terms of mathematical modelling. 
We have tested the possibility to determine
a state of the opinion with   data  issued from popularity and voting intentions polls 
available during the ``first tour''  of French presidential election of  April 2012. 
Of course the existence of such a state of the opinion 
remains an hypothesis, especially in the quantum case. 
We suggest that a possible further step is  to replace an ordered 
grading family of opinions by a  non-ordered set of political points of view.

\bigskip \bigskip    
 \noindent {\bf \large     Acknowledgments }  

\noindent 
The author thanks  Jerome R Busemeyer for his kind invitation to participate to  
 the session ``Quantum Decision Theory'' during the symposium
 {\it Foundations and Applications of Utility, Risk and Decision Theory} (FUR), 
Georgia State University, Atlanta (USA)  in July 2012. 

%

%

 \newpage 
\bigskip \bigskip  \bigskip 

\medskip

\end{document}